\documentstyle[aps]{revtex}

\input{psfig}
\begin{document}
%\baselineskip=0.5truecm
\draft
\preprint{}
\title{Differential dispersion relations with an arbitrary number 
of subtractions: a recursive approach}

\author{M.J. Menon$^{a}$, A.E. Motter$^{a}$ and B.M. Pimentel$^{b}$}

 \address{$ ^a$
Instituto de F\'{\i}sica `Gleb Wataghin'\\
Universidade Estadual de Campinas, Unicamp\\
13083-970 Campinas, SP, Brasil
}
 \address{$ ^b$
Instituto de F\'{\i}sica Te\'orica - UNESP\\
Rua Pamplona 145\\
01405-900 S\~ao Paulo, SP, Brasil 
}

%\date{\today}

\maketitle

\begin{abstract}
Making use of a recursive approach, derivative dispersion
relations are generalized for an arbitrary number of subtractions. The results for both cross even and odd amplitudes are theoretically consistent at sufficiently high energies and in the region of small momentum transfer.
\end{abstract}

\vskip 1.0truecm



e-mail: menon@ifi.unicamp.br, fax: +55-19-7885512, phone: +55-19-7885530

\vskip 1.0truecm

Dispersion-relation techniques have been widely used in the
investigation of high energy hadron scattering. In this area,
dispersion relations are rigorous consequences of local QFT
through the bridge represented by the concepts of analyticity and
polynomial boundedness \cite{teoremas}. This last condition means
that the scattering amplitude, as function of the center-of-mass
energy $\sqrt s$ and the four momentum transfer squared $t$, satisfies
the relation

\begin{equation}
\lim_{|s|\to \infty} f(s,t) \leq c|s|^k,
\end{equation}
for some finite $k$ and in all directions of the complex cut
$s$-plane \cite{teoremas,khuri}.

In the case of integral relations, it was proved that in the unphysical
region $0 < t < t_0$, where $t_0$ is a constant, the number of subtractions consistent with minimal conditions for the Froissart bound
is two \cite{jinmartin}. The same result was obtained in the physical region
(forward direction) without the use of the Froissart bound, but with the condition that the
forward-scattering amplitude does not become purely real at infinite
energies \cite{khurikinoshita}.

The limited usefulness of the integral dispersion relations, associated
with its non-local character, led to the introduction of
``quasi-local'' differential operators \cite{gribov,bks}. These derivative
dispersion relations (DDR) have played an important role in the
simultaneous investigation of total cross sections and the $\rho$-parameter
(ratio of the forward real and imaginary parts of the scattering amplitude),
for both $pp$ and $\overline p p$ elastic scattering \cite{matthiae}. 
Also, its mathematical aspects and range of validity have been extensively discussed
\cite{kolar} and, in the case of total hadronic cross sections, the smooth increase of this quantity with the energy assures its validity \cite{matthiae}.
Based on the general behaviour of the experimental data on total cross 
sections and the usual neglecting of the odd (crossing symmetric)
amplitude at the highest energies, the form generically used corresponds
to the first-order result for an even forward amplitude 
\cite{bks}

\begin{equation}
{Ref_{+}(s,t=0) \over s}= \frac{\pi}{2}\frac{d}{d\ln s}
\left[{Im f_{+}(s,t=0) \over s}\right],
\end{equation}
with the subtraction constant removed. This result is consistent with 
the usual convergence bounds inferred from the experimental data
presently available.

However, there is recent indication that total cross sections may
increase faster than usually expected at the highest energies 
\cite{niko,martinimenon}. It has also been shown that a possible failure
of polynomial boundedness (characteristic of some non-local approaches
and string theory), could be observed at sufficiently high energies
\cite{khuri}. Moreover, the odderon hypothesis (c-odd Regge trajectory) 
\cite{odderon} is supported by both perturbative QCD and fits to the 
experimental data \cite{gln} and a recipe to detect its signal at higher energies has been recently proposed \cite{gps}.

All these facts originate great expectations concerning the advent of the new generation of colliders, the Relativistic Heavy Ion Collider (RHIC) \cite{rhic} and the CERN Large Hadron Collider (LHC) \cite{lhc}. The reason is that energies never reached before in $pp$ collisions shall be investigated. It will also be possible to analyse the correct contribution of the c-odd amplitude and convergence properties of the amplitude with increasing energy, as represented by Eq. (1).

At this stage, even if we do not consider the extreme case of an effective
failure of the polynomial boundedness, it may be useful to investigate higher
bounds concerning convergence properties. Since the convergence of the integral
relations is controlled by subtractions and quasi-local connections (DDR) play
an important role in the analysis of experimental data, it may be suitable to extend a higher number of subtractions to the derivative relations.

To this end, in this work we first treat the singly subtracted derivative relation, introducting a recursive relation. Then we extend the $k$-subtracted relation to the derivative case, for both even and odd amplitudes.

We assume the usual cut structure in the complex energy plane \cite{bc,eden,collins}
and consider the high energy limit $s\gg m^{2}$, where $m$ is the proton mass, and the region where $|t|\lesssim m^{2}$. Taking into account the polynomial boundedness, Eq. (1), and neglecting the residues from the poles at the origin, the $k$-subtracted integral dispersion relation for even ($+$) and odd ($-$) crossing amplitudes reads \cite{formulation}

\begin{equation}
Re f_{\pm}(s,t)=
\frac{(s)^{k}}{\pi}
P\!\int_{s_{0}}^{+\infty} ds' 
\frac{Imf_{\pm}(s',t)}{(s')^{k}}
\left[\frac{1}{s'-s} \pm \frac{(-1)^{k}}{s'+s}\right].
\label{rdi}
\end{equation}

We first consider an even amplitude and one subtracted case ($k=1$). Following
Ref. \cite{bks}, after multiplying and dividing by $s^{\alpha}$ ($\alpha$ being a real parameter),
we integrate by parts and introduce the change of variable $s = e^{\xi}$, to
obtain

\begin{equation}
Re f_{+}(s,t)=\frac{s}{\pi}\int_{\ln s_{0}}^{\infty}d\xi ' s'^{\alpha -1}
\ln \coth\frac{1}{2}|\xi - \xi '|\left(\alpha -1 + \frac{d}{d\xi '}\right)
Im f_{+}(s',t)/s'^{\alpha}.
\end{equation}

In order to perform the integration we
shall take into account the following:
\begin{description}
\item ({\it i}) If $Im f_{+}(s',t)/s'^{\alpha}$ is an (real) analytic 
function of the variable $\xi'=\ln s'$ we can expand it in powers of 
$\xi ' -\xi$. We observe that this is not a trivial assumption.
\item ({\it ii}) It is easy to show that in the asymptotic  limits,

\begin{equation}
\lim_{|\xi '-\xi |\rightarrow 0}\ln \coth\frac{1}{2}|\xi '- \xi | = +\infty
, \:\:\: 
\lim_{|\xi '-\xi |\longrightarrow \infty}\ln\coth\frac{1}{2}|\xi '- \xi | = 
2e^{-|\xi ' -\xi |}\: ;
\label{28}
\end{equation}
\item ({\it iii}) Since $s\gg s_{0}$, the result ({\it ii}) 
means that for large $|s' - s|$ the contribution to the integral is
small. In this case, we can take $s_{0}\rightarrow 0$ so that  $\ln s_{0}
\rightarrow -\infty$, as the lower limit in Eq. (4).
\end{description}

From ({\it i}) and ({\it iii}) and assuming that the series may be
integrated term by term we obtain

\begin{equation}
Ref_{+}(s,t)=s^{\alpha}\sum_{n=0}^{\infty}\frac{d^{(n)}}{d \ln s^{(n)}}
\left( Im f_{+}(s,t)/s^{\alpha}\right)
\frac{I_{n}}{n!}\:\: ,
\label{29}
\end{equation}
where $I_{n}$ represents the integral in the variable $\xi '$ ,

\begin{equation}
I_{n}=\frac{1}{\pi}\int_{-\infty}^{+\infty}d\xi 'e^{(\alpha -1)(\xi '-\xi)}
\ln\coth \frac{1}{2}|\xi ' -\xi|\left(\alpha -1+\frac{d}{d\xi '}\right)
(\xi '-\xi)^{n}.
\label{30}
\end{equation}
Denoting $\xi ' - \xi\equiv y$ and through integration by parts, this equation may be put in the form

\begin{equation}
I_{n}=\frac{1}{\pi} \ln\coth\frac{1}{2}|y|\:\: e^{(\alpha -1)y}y^{n}
{\large \mid}_{-\infty}^{+\infty}
+ \frac{1}{\pi}\int_{-\infty}^{+\infty}dy \frac{e^{(\alpha -1)y}}{\sinh y}y^{n}.
\label{32}
\end{equation}

From the first term it is easy to see that convergence to a finite value
(zero in this case) is obtained
only for limited values of the parameter $\alpha$, namely,
$
\alpha \in (0,2).
$ 
For $n=0$ the above integral may be evaluated in the complex plane,
$
I_{0}= \tan \left(\frac{\pi}{2}(\alpha -1)\right) .
$
For $n=1,2,3,...$, the integral $I_{n}$ may be formally determined if we differentiate
the second term in Eq. (8), taking $\alpha$ as variable and changing the order of integration
and differentiation. This is allowed since the integrand is continuous and the
integral is uniformly convergent.
With this we obtain a novel recursive relation

\begin{equation}
I_{n}=\frac{d^{(n)}I_{0}}{d\alpha^{(n)}}=
\frac{d^{(n)}}{d\alpha^{(n)}}\tan (\frac{\pi}{2}(\alpha -1)).
\label{36}
\end{equation}
Finally, each term of the series in Eq. (\ref{29}) may be obtained through
recursive derivation. 
Moreover, the result may be put in a closed
form which is suitable for theoretical manipulations: Substituting Eq. (\ref{36})
into Eq. (\ref{29}) with the tangent expressed by its series
we get

\begin{equation}
Ref_{+}(s,t)=s^{\alpha}\tan 
\left[ \frac{\pi}{2}\left(\alpha -1 +\frac{d}{d\ln s}\right) \right]
Imf_{+}(s,t)/s^{\alpha}\: .
\label{40}
\end{equation}
The above expression connects the real part of an even amplitude with
the derivatives of the imaginary part at the same energy. Due to the derivative character involved, it shall be considered as a quasi-local relation.

In what follows we generalize this result for an arbitrary number of 
subtractions and also for odd amplitudes. To this aim we begin with the integral relation for an arbitrary number of subtractions. From Eq. (\ref{rdi}), for $k=2n$ and $k=2n-1$ we have for 
the even amplitudes

\begin{equation}
Ref_{+}(s,t) = \frac{s^{2n-1}}{\pi} P\!\int_{s_{0}}^{\infty}
ds' Im\left(\frac{f_{+}(s',t)}{s'^{2n-1}}\right)
\left[\frac{2s}{s'^{2}-s^{2}}\right].
\label{41}
\end{equation}
Analogously, for $k=2n$ and $k=2n+1$ the odd amplitude reads

\begin{equation}
Ref_{-}(s,t) = \frac{s^{2n-1}}{\pi} P\!\int_{s_{0}}^{\infty}
ds' Im\left(\frac{f_{-}(s',t)}{s'^{2n-1}}\right)
\left[\frac{2s^{2}}{s'(s'^{2}-s^{2})}\right].
\label{42}
\end{equation}
Defining
\begin{equation}
g_{+} (s',t) = \frac{f_{+}(s',t)}{s'^{2(n-1)}}
\end{equation}
and from Eq. (3) with $k=1$ and (11), it follows that 
$g_{+}$ satisfies a two-subtracted integral dispersion relation for an even
function. Although $g_{+}$ might have a pole at the origin, this does not bring
any disagreement with the approach as can easily be
verified. In addition, if we assume assertion $(i)$, $g_{+}$ verifies all the necessary conditions for the 
derivative relation  (10), so that

\begin{equation}
Ref_{+}(s,t)=s^{2(n-1)+\alpha}
\tan\left[\frac{\pi}{2}\left( \alpha -1 +\frac{d}{d\ln s}\right) \right]
Im f_{+}(s,t)/s^{2(n-1)+\alpha}.
\label{44}
\end{equation}
Analogously, defining 
\begin{equation}
g_{+}(s',t)=\frac{f_{-}(s',t)}{s'^{2n-1}} 
\end{equation}
the same arguments lead to 

\begin{equation}
Ref_{-}(s,t) = s^{2n-1+\alpha}
\tan\left[\frac{\pi}{2}\left( \alpha -1 +\frac{d}{d\ln s}\right) \right]
Im f_{-}(s,t)/s^{2n-1+\alpha}.
\label{46}
\end{equation}
Differently from the even parity, this expression shows that the
1st  and 2nd subtractions are quite distinct in the odd case.
\vspace{0.3cm}

The essential results of this work are:
(a) introduction of a recursive relation in the parameter $\alpha$,
Eq. (9), in order to obtain a compact form for the differential  relation, Eq. (10);
(b) generalization of the  derivative relations for an arbitrary number of subtractions,
for both even and odd amplitudes, Eqs (14) and (16) respectively, 
near the forward scattering.
\vspace{0.1cm}

M.J.M and B.M.P. are thankful to CNPq and A.E.M. to Fapesp for
financial support.

\end{document}